\documentclass{aa}  

\usepackage{graphicx}
\usepackage{txfonts}
\usepackage{color}
\makeatletter
\newcommand{\change}[2][\@empty]{{\color{blue} #2}%
 \ifx#1\@empty\else{\color{green}\footnote{%
     \ifx#2{instead of}\else{removed}\fi: \color{green}\normalsize#1}}\fi}
\makeatother

\begin{document}

   \title{On the speed and acceleration of electron beams triggering interplanetary type III radio bursts}
 \titlerunning{Electron beams}

   \author{V. Krupar
          \inst{1}
          \and
					E. P. Kontar
          \inst{2}
          \and
					J. Soucek
          \inst{1}
          \and
					O. Santolik
          \inst{1,3}
          \and
					M. Maksimovic
          \inst{4}
          \and
					O. Kruparova
          \inst{1}
          }

   \institute{Institute of Atmospheric Physics, The Czech Academy of Sciences, Prague 14131, Czech Republic\\
              \email{vk@ufa.cas.cz}
         \and
             School of Physics and Astronomy, University of Glasgow, Glasgow G12 8QQ, UK
         \and
             Faculty of Mathematics and Physics, Charles University, Prague 121 16, Czech Republic
         \and
             LESIA, UMR CNRS 8109, Observatoire de Paris, Meudon 92195, France
             }


 
  \abstract
   {}
   {Type III radio bursts are intense radio emissions triggered by beams of energetic electrons often associated
with solar flares. These exciter beams propagate outwards from the Sun along an open magnetic field line in the corona
and in the interplanetary (IP) medium.}
   {We performed a statistical survey of 29 simple and isolated IP type III bursts
observed by STEREO/Waves instruments between January 2013 and September 2014.
We investigated their time-frequency profiles in order to derive the speed and acceleration of exciter electron beams.}
   {We show these beams noticeably decelerate in the IP medium. Obtained speeds range from $\sim$~0.02c
	up to $\sim$~0.35c 
depending on initial assumptions. It corresponds to electron energies between tens of eV and hundreds of keV, and in order to explain
the characteristic energies or speeds of type III electrons ($\sim 0.1$c) observed simultaneously with Langmuir waves at 1~au,
the emission of type III bursts near the peak should be predominately at double plasma frequency.
Derived properties of electron beams can be used as input parameters for computer
simulations of interactions between the beam and the plasma in the IP medium.}
   {}


   \maketitle
%

\section{Introduction}

Type III radio bursts are produced by beams of suprathermal electrons
accelerated near the Sun's surface during solar flares \citep{1950AuSRA...3..541W}.
These exciter beams propagate outwards from the Sun along an open magnetic field line in the corona
and the interplanetary (IP) medium at large distances beyond $1$~au, where suprathermal electrons
can be detected in situ by spacecraft. They locally produce a bump-on-tail instability that leads to a generation
of electrostatic Langmuir waves near the local plasma frequency $f_{\rm{pe}}$.
The conversion of Langmuir waves into escaping electromagnetic emission is believed to be due to
nonlinear interactions involving electromagnetic waves: type III radio bursts near $f_{\rm{pe}}$ (the fundamental, \textit{F}-component)
or near $2f_{\rm{pe}}$ (the harmonic, \textit{H}-component), which can be observed
remotely \citep{1970SoPh...12..266L,1976Sci...194.1159G}. This conversion involves the wave-wave interactions of Langmuir,
ion acoustic and electromagnetic waves \citep{1958SvA.....2..653G}.
Since electron beams propagate away from the Sun, the characteristic emission frequency decreases
to consecutively lower frequencies corresponding to a decrease in $f_{\rm{pe}}$.

Although we can distinguish the \textit{F}- and \textit{H}-components from metric to decametric wavelengths
(\textit{i.e.} of coronal origin), it is difficult to determine the
observed component of type III bursts at longer wavelengths, which are generated in the IP medium. However, we can recognize
the \textit{F}- and \textit{H}-components for cases where type III triggering electron beams intersect
the spacecraft, those associated with Langmuir waves observed \textit{in situ} \citep{1980ApJ...236..696K}.
Nonetheless, such events occur rather seldom. Although IP type III bursts are generally
assumed to be emitted at the harmonic of the electron plasma frequency \citep{1992ApJ...394..340R,1999A&A...348..614M},
we will discuss presence of both components separately.

Since early observations, the characteristic exciter speeds associated with type III bursts have
normally been found in the range $0.1$~--~$0.6$c.
\citet{1987A&A...173..366D} investigated both onset and peak times of type III bursts associated with
measurements of Langmuir waves and suprathermal electrons by the ISEE-3 spacecraft
assuming presence of the \textit{F}-component alone \citep{1978ITGE...16..191S,1978ITGE...16..153A}.
They showed the exciter beam speed of $0.14$c and $0.07$c using onset and peak times, respectively.
\citet{1994A&A...289..957H} have studied onset times of the type III bursts observed by the URAP instrument onboard the Ulysses spacecraft \citep{1992A&AS...92..291S}
using the same approach as \citet{1987A&A...173..366D}.
\citet{1994A&A...289..957H} concluded that the exciter beam speed ranges from $0.04$c to $0.13$c.
However, the variation in these speeds proved to be more challenging.
\citet{1970SoPh...15..433F} have suggested that the beams are propagating over 1~au with little
deceleration.
Later, \citet{1996AIPC..382...62P} suggest that deceleration occurs before about 0.1~au,
and that there is no significant correlation between the initial and final speeds.
The STEREO provides radio data with significantly better resolution than used in the above-mentioned studies.

In this paper, we present statistical results on speed and acceleration of exciter beams of
IP type III radio bursts observed by the Solar TErrestrial
RElations Observatory (STEREO).
In Section~\ref{S-STEREO} we focus on the instrumentation.
In Section~\ref{S-dataset} we describe our data set.
Then we show a detailed analysis of one IP type III radio burst to illustrate the methods we used (Section~\ref{S-event}).
We present statistical results for exciter speeds and acceleration in Section~\ref{S-Results}.


\section{Observations and data analysis}

      \label{S-data}

\subsection{The STEREO spacecraft}
\label{S-STEREO}
STEREO consists of two identical spacecraft orbiting the Sun and providing a unique
stereoscopic view of the solar-terrestrial system \citep{2008SSRv..136....5K}.
The STEREO/Waves instrument records electric field fluctuations by three monopole antennas \citep{2008SSRv..136..487B,2008SSRv..136..529B}.
For analysis of type III radio bursts we use data recorded by the Low Frequency Receiver (LFR, $10$~kHz~--~$160$~kHz in $32$~logarithmically spaced frequency channels)
and High Frequency Receiver (HFR, $125$~kHz~--~$16.025$~MHz in $319$~linearly spaced frequency channels).
The time resolution of both receivers is $38$~seconds.
With LFR and HFR we can investigate radio emissions that originate somewhere from $2$~$\rm{R_S}$ up to the $1$~au of the Sun
when assuming an average electron density model in the solar wind \citep{2008SSRv..136..549C}.
However, we are unable to accurately calculate frequency drifts of type III radio bursts above $1$~MHz
owing to low time resolution of the STEREO/\textit{Waves} instrument.
Therefore we limit our analysis to below $1$~MHz.
Both receivers can operate either in the three-monopole mode or in the dipole-monopole mode.
We analysed the data only in the three-monopole mode. The LFR and HFR have been operating
in this mode since January 2013 and May 2007, respectively.
We subtracted receiver background levels from the data to improve the signal-to-noise ratio.
These levels were calculated as median values over a long period of 15 days for each channel--antenna configuration separately.
More details on method used can be found in \citet{2010AIPC.1216..284K,2012JGRA..11706101K,2014SoPh..289.3121K}.

\subsection{Data set}
\label{S-dataset}
We performed a statistical analysis of 29~type III radio bursts observed by STEREO/\textsl{Waves}
between January 2013 and September 2014.
Only intense, simple, and isolated cases (\textit{i.e.} without overlapping) with a low frequency cutoff below $100$~kHz have been included in our data set.
These constraints resulted in relatively few suitable events for time -- frequency investigation.
However, this helped to select radio bursts a with smooth monotone relation between peak time and frequency.
\subsection{September 27, 2013 type III radio burst}
\label{S-event}

 \begin{figure}
 \centering
 \includegraphics[width=0.50\textwidth]{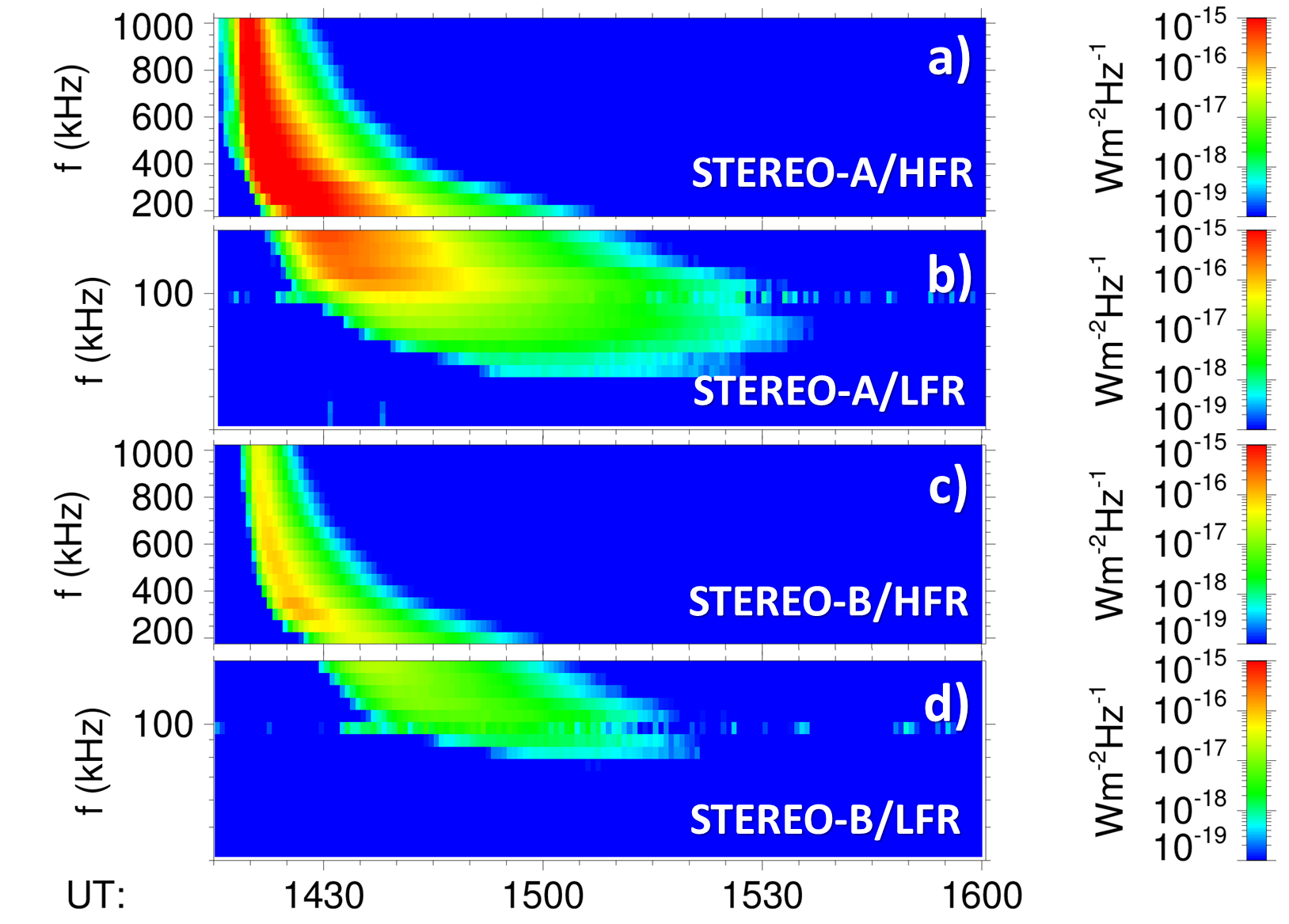}
 \caption{Analysis of measurements recorded from 14:15 to 16:00 UT on September 27, 2013: the flux density $S$ for STEREO-A (panels a, and b)
and STEREO-B (panels c, and d).}
              \label{fig:20130927_psd}%
 \end{figure}

We show an analysis of an IP type III radio burst on September 27, 2013 to illustrate the methods used.
Figures~\ref{fig:20130927_psd}a~--~\ref{fig:20130927_psd}d display the flux density $S$ measured by LFR and HFR
onboard STEREO-A and STEREO-B. Both spacecraft detected the simple and isolated type III radio burst with a starting time
of about 14:20 UT. The low frequency cutoff is $60$~kHz and $80$~kHz for STEREO-A and STEREO-B, respectively.
We excluded frequencies above $1$~MHz because of the insufficient time resolution of the receiver.
During this event STEREO-A was $140^\circ$ west of a Sun-Earth line at
$0.97$~au from the Sun, whereas STEREO-B was $146^\circ$ east and $1.05$~au of the
Sun.

We investigated times corresponding to peak fluxes since onset times are difficult to accurately identify owing to
density variations that result in fluctuations of the quasi-thermal noise \citep{1989JGR....94.2405M}.
Frequency channels have been converted to radial distances from the Sun
considering both the \textit{F}-component ($r_{\rm{F}}$) and the \textit{H}-component ($r_{\rm{H}}$)
using the electron density model in the IP medium by \citet{1999ApJ...523..812S}.
It allows the obtained times to be corrected
for a light travel time frequency by frequency from the source to the spacecraft with a radial distance from the Sun ($r_{\rm{S/C}}$):
\begin{equation}  \label{eq:correct}
t_i(f)~=~t(f)-\frac{r_{\rm{S/C}}-r_i(f)}{c}\mbox{,}
\end{equation}
where $f$ represents frequency, c is a speed of light, and $i$ stands for the \textit{F}- and \textit{H}-components.
This correction assumes that radio sources are located on the Sun-to-spacecraft line.
The validity of this assumption has been tested using computer simulations in Appendix A.

Previous studies of exciter beam speed were dedicated to events associated Langmuir waves observed
in situ \citep{1987A&A...173..366D,1994A&A...289..957H,1994SoPh..154..335R} resulting in
different assumptions on radio source locations. Plasma parameters
measured onboard were used to estimate the Parker spiral geometry together with the density profile along it \citep{1958ApJ...128..677P}.
However, this approach does not consider a propagation of plasma parameters between the Sun, so the spacecraft located at $1$~au.
It can take up to several days for the solar wind to travel between the corona and the spacecraft and both plasma speed and density,
used for the Parker spiral reconstruction, may differ significantly from measurements in situ.
Because of this drawback and since our events were not associated with Langmuir waves, we assume that the radio sources lies on the
Sun-to-spacecraft line, so the average electron density model in the ecliptic was used.

A comparison between corrected times and frequencies converted into radial distances allows speed
and acceleration of exciter beams to be derived.
However, this simple method does not take several important ambiguities into account.
The first is that we are unable to distinguish the \textit{F}- and \textit{H}-components of IP type III radio bursts.
The distance $r_{\rm{H}}$ is typically twice greater than $r_{\rm{F}}$ resulting in about twice greater beam
speed and acceleration for the \textit{H}-component than for the \textit{F}-component.
Therefore we discuss our results considering the presence of both components throughout the paper.
Moreover, triangulated radio sources of IP type III radio bursts lie at considerably
(five times for the \textit{F}-component and three times for the \textit{H}-component)
larger radial distances from the Sun than the electron density models predict \citep{1984A&A...140...39S,2014SoPh..289.4633K}.
It indicates that we observed only scattered images of real sources that are apparently very extended.
We thus conclude that the presented method provides us with a lower bound estimate of the exciter speed.

 \begin{figure}
 \centering
 \includegraphics[width=0.50\textwidth]{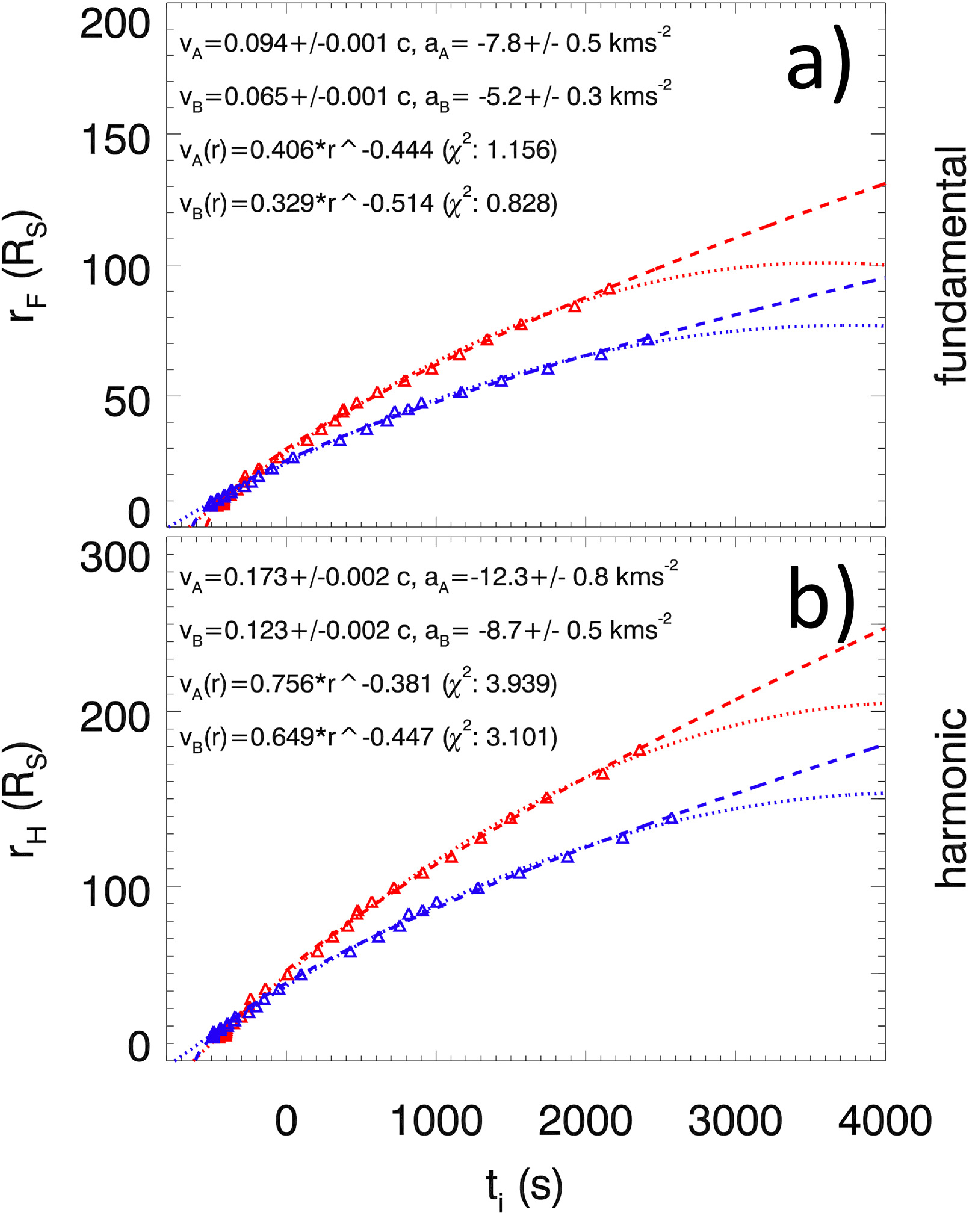}
 \caption{Analysis of measurements recorded from 14:15 to 16:00 UT on September 27, 2013: radial distances from the Sun
\textit{vs.} corrected times for STEREO-A (in red) and STEREO-B (in blue) assuming
the \textit{F}- (panel a) and \textit{H}-component (panel b).
Dotted lines represent results of the polynomial model (equation \ref{eq:linear}). Dashed lines show the power law model
(equation \ref{eq:power}). Obtained parameters of both models are shown on the top.}
              \label{fig:20130927_line}%
 \end{figure}

Figures \ref{fig:20130927_line}a and \ref{fig:20130927_line}b display
radial distances from the Sun ($r_{\rm{F}}$ and $r_{\rm{H}}$) as a function of
corrected times ($t_{\rm{F}}$ and $t_{\rm{H}}$).
We used two models to fit the $r_{\rm{i}}$ \textit{vs.} $t_{\rm{i}}$ dependence. Initially, we considered
a constant acceleration, when the speed ($v_{\rm{i}}$) is a linear function of time ($t_{\rm{i}}$):
\begin{equation}  \label{eq:linv}
v_{\rm{i}}(t_{\rm{i}})~=~v_{\rm{0i}}+a_{\rm{i}} t_{\rm{i}}\mbox{,}
\end{equation}
where $v_{\rm{0i}}$ is a beam speed at $t_{\rm{i}}=0$ (an initial speed in [$\rm{kms^{-1}}$]), and $a_{\rm{i}}$ represents the beam acceleration in [$\rm{kms^{-2}}$].
The corresponding relation between $r_{\rm{i}}$ and $t_{\rm{i}}$ is then
\begin{equation}  \label{eq:linear}
r_{\rm{i}}(t_{\rm{i}})~=~r_{i0}+v_{\rm{0i}}t_{\rm{i}}+\frac{1}{2}a_{\rm{i}} t_{\rm{i}}^2\mbox{,}
\end{equation}
where $r_{i0}$ is an initial distance from the Sun at $t_{\rm{i}}=0$. We denote it as "the polynomial model" which
represents the simplest approach for estimation of both initial speed and acceleration.

The second model assumes that a beam speed $v_{\rm{i}}$ has a power law dependence on the radial distance $r_{\rm{i}}$ from the Sun as
\begin{equation}  \label{eq:varb}
v_{\rm{i}}(r_{\rm{i}})~=~\alpha_{\rm{i}} (r_{\rm{i}}/\rm{R_S})^{\beta_{\rm{i}}}\mbox{,}
\end{equation}
where $\alpha_{\rm{i}}$ represents the beam speed at $1~\rm{R_S}$ (a power law initial speed) and $\beta_{\rm{i}}$ denotes power-law index, so that 
positive values of $\beta_{\rm{i}}$ represent acceleration.
Since we measure $r_{\rm{i}}$ \textit{vs.} $t_{\rm{i}}$ we need to solve the first-order nonlinear differential equation:
\begin{equation}  \label{eq:varb_r}
\frac{dr_{\rm{i}}}{dt_{\rm{i}}}~=~\alpha_{\rm{i}} (r_{\rm{i}}/\rm{R_S})^{\beta_{\rm{i}}}\mbox{.}
\end{equation}
By separating variables we obtain the following relation between $r_{\rm{i}}$ and $t_{\rm{i}}$:
\begin{equation}  \label{eq:power}
r_{\rm{i}}(t_{\rm{i}})~=~[(1-\beta_{\rm{i}})(\alpha_{\rm{i}} t_{\rm{i}} + C_{\rm{i}})\rm{R_S}^{-\beta_{\rm{i}}}]^{\frac{1}{1-\beta_{\rm{i}}}}\mbox{,}
\end{equation}
where $C_{\rm{i}}$ is the constant of integration.
Since both numerical simulations \citep{2001SoPh..202..131K} and 
theoretical relationships suggest that the velocity should vary as a power of the distance of the source from the Sun \citep{1992SoPh..137..307R}, 
we denote it as "the power law model". The acceleration $a_{\rm{i}}$ for this model can be obtained by differentiation of
equation (4):
\begin{equation}  \label{eq:power_a}
a_{\rm{i}}(r_{\rm{i}})~\equiv~\frac{dv_{\rm{i}}}{dt}~=~\beta_{\rm{i}}\alpha_{\rm{i}}\left(\frac{r_{\rm{i}}}{{\rm{R_S}}}\right)^
{\beta_{\rm{i}}-1}\frac{1}{{\rm{R_S}}}\frac{dr_{\rm{i}}}{dt}~=~
\frac{\beta_{\rm{i}}\alpha^2_{\rm{i}}}{\rm{R_S}}\left(\frac{r_{\rm{i}}}{{\rm{R_S}}}\right)^{2\beta_{\rm{i}}-1}\mbox{.}
\end{equation}

We performed a polynomial regression of the second order to estimate type III exciter beam parameters according to
the polynomial model (equation \ref{eq:linear}).
Calculated initial speeds are greater at STEREO-A than at STEREO-B. This difference
is probably related to the time correction method. The stronger signal at STEREO-A
suggests that the electron beam propagates more towards STEREO-A than STEREO-B.
The time correction method thus underestimates the calculated speed slightly at the latter (Appendix A).
The \textit{F}-component assumption (panel a) provides us with slower
speeds when compared to the \textit{H}-component as it can be expected.
Our results indicate that the electron beam decelerates ($a_{\rm{F}}\sim -7$~$\rm{km~s^{-2}}$ and $a_{\rm{H}}\sim~-12$~$\rm{km~s^{-2}}$)
depending on an assumption on the observed component.
For the power law model, we used a nonlinear least squares fit of the spacecraft data with equation (\ref{eq:power}).
The parameters of the model obtained with a $\chi^2$ goodness-of-fit statistic weighted by the measurement error are shown in
Figure \ref{fig:20130927_line}.
Negative values of $\beta$ confirm that the electron beam decelerates.

Both models fit the data very well.
We observe a discrepancy between the two models for larger distances
from the Sun when the polynomial model predicts greater deceleration when compared to the power model.
Since exciter beams typically travel far beyond $1$~au, the exciter power law model seems to be more applicable for larger
distances from the Sun.
However, exciter speeds farther from the Sun, and the initial speeds ($v_{i0}$, $\alpha_{\rm{i}}$),
where we do not have any data, are dubious since they represent an extrapolation of both models.

\subsection{Statistical results}
\label{S-Results}

\begin{figure*}
\centering
\includegraphics[width=0.70\textwidth,angle=0]{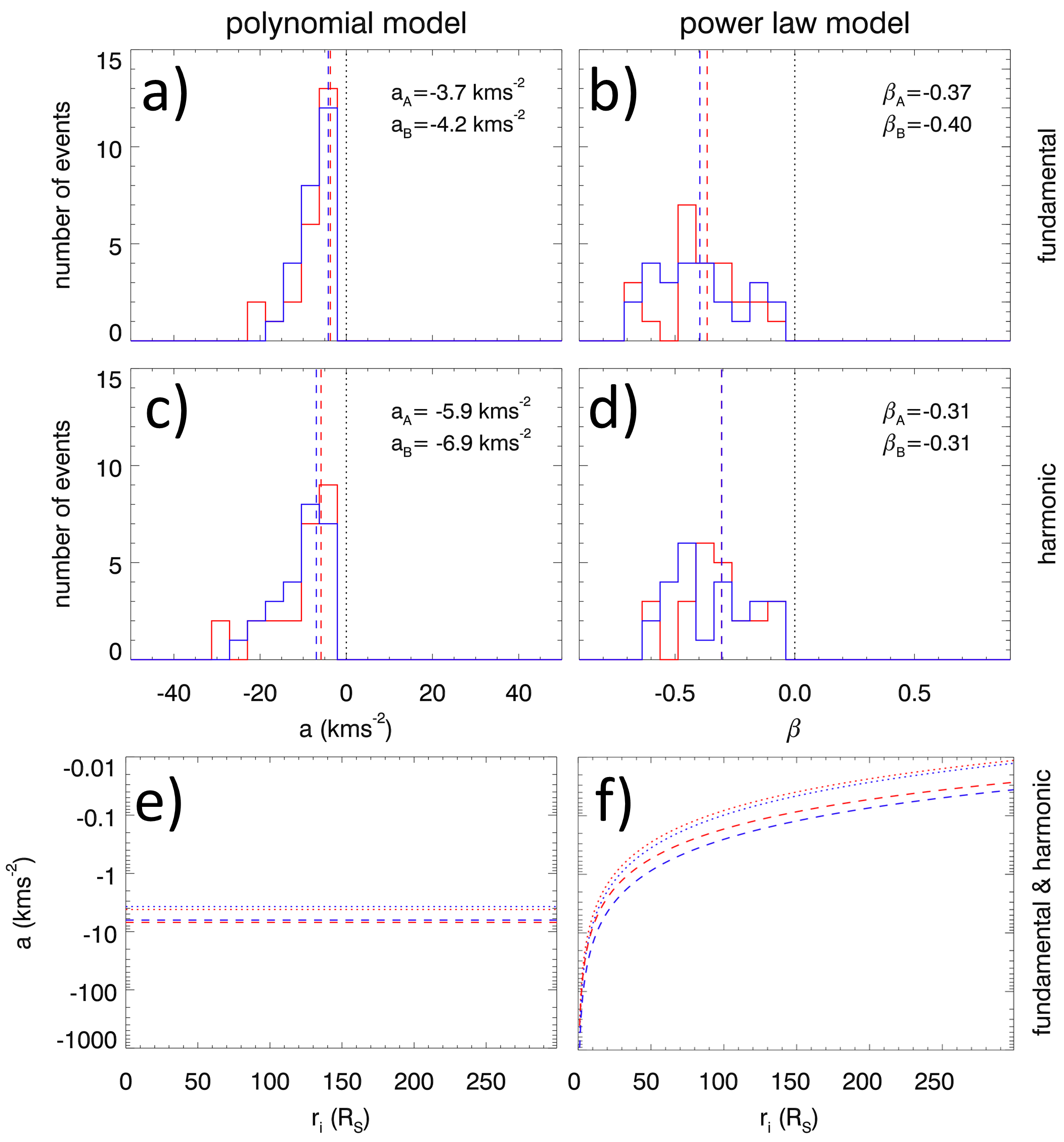}
\caption{Panels a -- d: Histograms of the acceleration coefficient $a$ (the polynomial model, equation \ref{eq:linear})
and the acceleration coefficient $\beta$ (the power law model, equation \ref{eq:power}) assuming the \textit{F}- and \textit{H}-component
for STEREO-A (in red) and STEREO-B (in blue). Dashed lines indicate median values. Black dotted lines show zero values. Panels e \& f:
Acceleration vs. distances for the polynomial and power-law models for STEREO-A (in red) and STEREO-B (in blue). Dotted and dashed lines
represent \textit{F}- and \textit{H}-components, respectively.}
\label{fig:histo_acc}
\end{figure*}

We performed the above described propagation analysis on 29~IP type III radio bursts
observed by STEREO/\textit{Waves} between January 2013 and September 2014
case by case (Section 2.2).

Figures \ref{fig:histo_acc}a -- \ref{fig:histo_acc}d display histograms of the acceleration coefficient $a_{\rm{i}}$
and $\beta_{\rm{i}}$.
Median values of the acceleration for the \textit{F-}component $a_{\rm{F}}$ are $\sim$~$-4$~km~s$^{-2}$
while the acceleration for the \textit{H}-component $a_{\rm{H}}$ is about $50\,\%$ larger than $a_{\rm{F}}$.
However, both $\beta_{\rm{F}}$ and $\beta_{\rm{H}}$ are comparable for the power law model ($\sim~-0.35$).
Figures \ref{fig:histo_acc}e \& \ref{fig:histo_acc}f show the relation between the acceleration and the radial distance from the Sun
for both models using median values from panels above. While the acceleration is obviously constant
for the polynomial model (Figure \ref{fig:histo_acc}e), it decreases for the power law model with a radial distance (Figure \ref{fig:histo_acc}f).
The acceleration for both models are comparable between $5$~$\rm{R_S}$ and $20$~$\rm{R_S}$.

\begin{figure*}
\centering
\includegraphics[width=0.70\textwidth,angle=0]{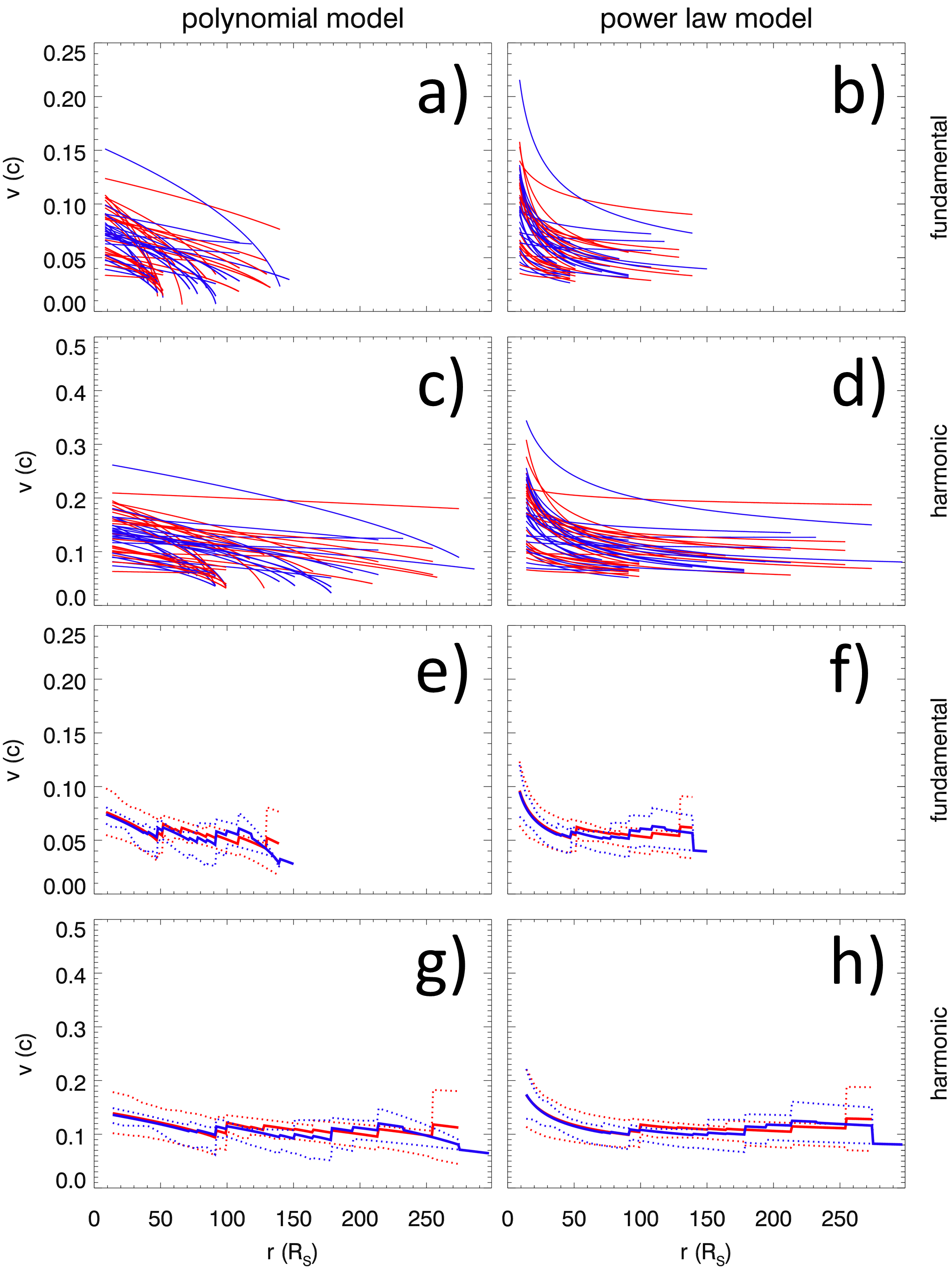}
\caption{Panels a, b, c, and d: Exciter beam speeds $v_{\rm{i}}$ \textit{vs.} radial distance $r_{\rm{i}}$ for all events considering
the polynomial model (equation \ref{eq:linear}) and the power law model (equation \ref{eq:power})
for STEREO-A (in red) and STEREO-B (in blue).
Panels e, f, g, and h: Median values (solid lines) and $25$th/$75$th percentiles (dotted lines) of data in panels
a, b, c, and d.}
\label{fig:speed_dist}
\end{figure*}

Figure \ref{fig:speed_dist} shows
results on exciter beam speeds $v_{\rm{i}}$ vs. radial distance $r_{\rm{i}}$ for all events.
Curves represent only the data that have been measured by STEREO (\textit{i.e.} above the low frequency cutoff case by case,
Figures \ref{fig:speed_dist}a -- \ref{fig:speed_dist}d).
Calculated speeds range from $\sim~0.02$c to $\sim~0.22$c and from $\sim~0.03$c
to $\sim~0.35$c for
the \textit{F}- and \textit{H}-components, respectively.
From individual kinematic curves, we calculated median beam speeds with the 25th and the 75th percentiles
(Figures \ref{fig:speed_dist}e -- \ref{fig:speed_dist}h). The statistical curves we obtained contain jumps due to the termination of some bursts.
Median values of the speeds decrease from $\sim~0.09$c to $\sim~0.04$c and from $\sim~0.16$c to $\sim~0.09$c
for the \textit{F}- and \textit{H}-component, respectively.
It suggests that the exciter beams of IP type III radio bursts decelerate in the solar wind.
However, other interpretations of this statistical results can be proposed: the observed deceleration could be an apparent observational effect caused by stronger scattering and thus larger refractive index of radio waves at lower frequencies.
The important role of scattering of IP type III bursts has been demonstrated via the Monte Carlo
simulations by \citet{2007ApJ...671..894T}.
The power law model provides us with more reasonable beam speeds when extrapolated to $1$~au,
while the polynomial model predicts a beam halt before $1$~au which is not generally observed.

%
  
%
%

\section{Conclusions}
\label{S-Conclusions}

We investigated the properties of 29~simple and isolated type III radio
bursts observed by the two STEREO spacecraft.
The \textit{F}- and \textit{H}-components have been considered
using two different models for estimating the exciter beam speed and acceleration.
We show a detailed analysis of one event from our data set as an example.
The presented method provides us with a lower bound estimate of the exciter speed.

Our results suggest that exciter beams decelerate in the IP medium (Figure \ref{fig:histo_acc}).
The collisional losses cannot account for this effect: the collisional mean free path of electron
with kinetic energy $\sim10$~keV in typical solar wind plasma is higher than $1$~au and the cumulative effect of collisions
on the beam energy is negligible. The deceleration can be explained by the energy loss of electron beams due to instabilities
producing Langmuir waves and the effect of solar-wind inhomogeneities on these waves. This loss of the energy is a two-step process. 
First, Langmuir waves are generated by the front  the beam. Second, these wave are refracted in decreasing plasma
density towards high wave numbers (lower phase velocities), thereby reducing the level of waves that
can be re-absorbed by the beam \citep[e.g.][]{2001SoPh..202..131K}. Large scale simulations of electron transport
show that a large part of the initial energetic electron energy
is lost via Langmuir waves due to plasma inhomogeneity \citep{2013SoPh..285..217R}.
The obtained acceleration $a_{\rm{i}}$ ranges from $\sim~-4$~$\rm{km~s^{-2}}$ to $\sim~-7$~$\rm{km~s^{-2}}$.
Derived exciter speeds vary between $\sim~0.02$c and $\sim \change~0.35$c, which corresponds to electron energies
from tens of eV up to hundreds of keV (Figure \ref{fig:speed_dist}), where the higher speeds correspond to the \textit{H}-component.
A comparison with energies of electron beams detected in situ suggest that
the \textit{H}-component predominantly occur in our data set \citep{1996GeoRL..23.1211L}.
However, the deceleration could also be related to stronger scattering at lower frequencies.
Presented properties of type III exciter beams can be used as input parameters
for computer simulations of beam -- plasma interactions in the IP medium.


\appendix

\section{Time correction method validation}\label{app:time}

 \begin{figure}
 \centering
 \includegraphics[width=0.50\textwidth]{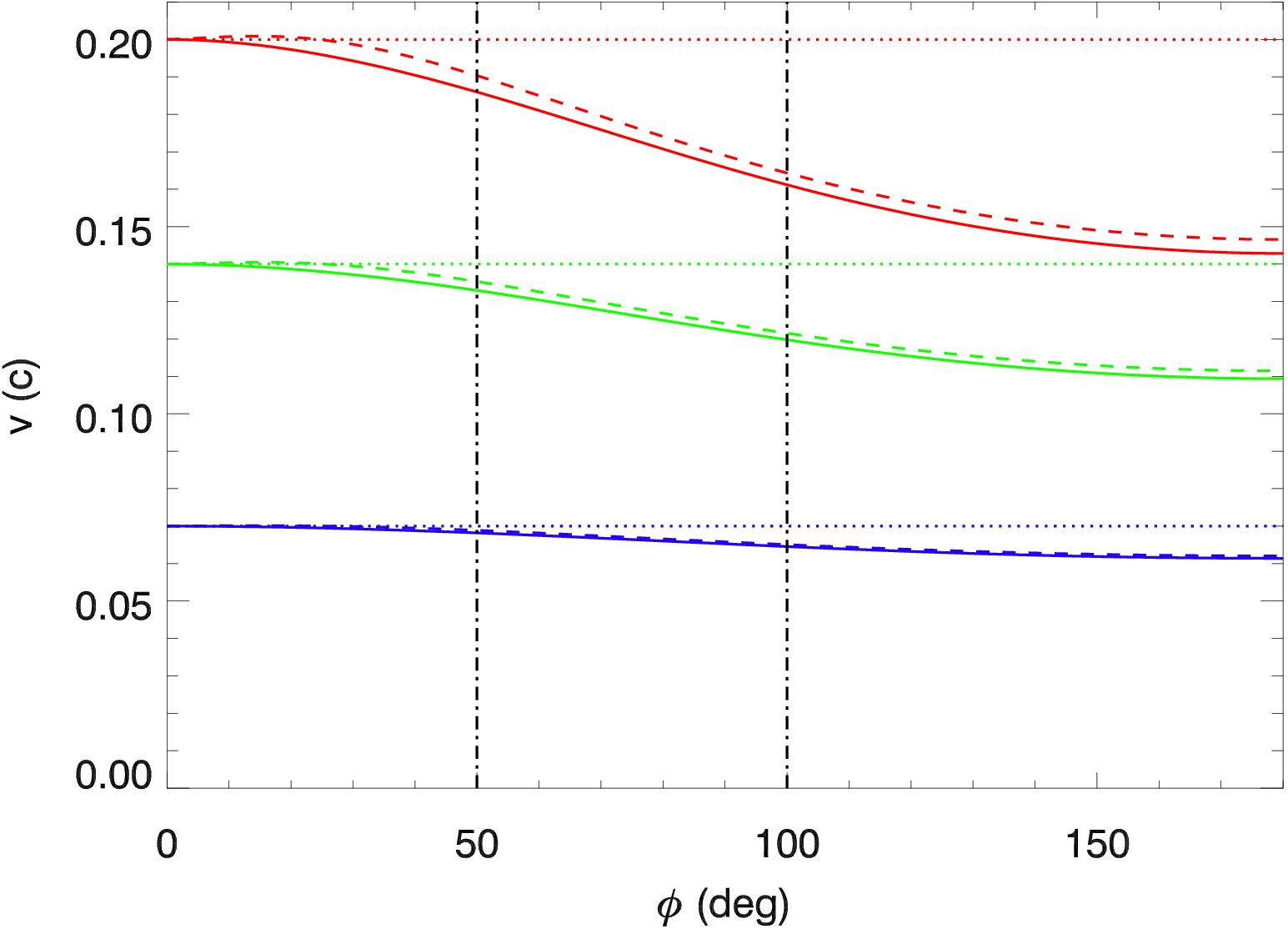}
 \caption{Simulated beam speeds \textit{vs.} deviation angle $\Delta\phi$. Colors denote beam speeds (red: $0.2$c, green: $0.14$c, and blue: $0.07$c).
The fundamental and harmonic components are represented by solid and dashed lines, respectively. Horizontal lines indicate initial speeds.
Dash-dotted lines show maximal and mean separation angles between the two STEREO in our data set.}
              \label{fig:sim}
 \end{figure}

We validated our time correction method via simulations of electron beams propagating away from the Sun (Figure \ref{fig:sim}).
Simulated beams travel with a given constant speed along the X axis while emitting radio emissions at frequencies between $40$~kHz and $1$~MHz
(\textit{i.e.} at a frequency range of STEREO/Waves used in this study).
We have considered the \textit{F-} and \textit{H-}components using the electron density model by \citet{1999ApJ...523..812S}.
A modeled spacecraft lies on a circle around the Sun with a radius of $1$~au in the ecliptic (Z=0), so its position can be
parametrized by angle $\phi=\arctan{\frac{Y}{X}}$.
We applied the time correction method (Eq. \ref{eq:correct}) to observed times by the modeled spacecraft.
Our results show that differences between initial and derived beam speeds are negligible for small angles.
For a mean separation angle between the two STEREO spacecraft in our data set ($\phi=50^\circ$), this difference ranges from $1$\,\% to $7$\,\%.
The difference at $\phi=100^\circ$ varies between $7$\,\% and $20$\,\%.
Moreover, variations between the \textit{F-} and \textit{H-}components are minor.

 \begin{figure*}
 \centering
 \includegraphics[width=0.70\textwidth]{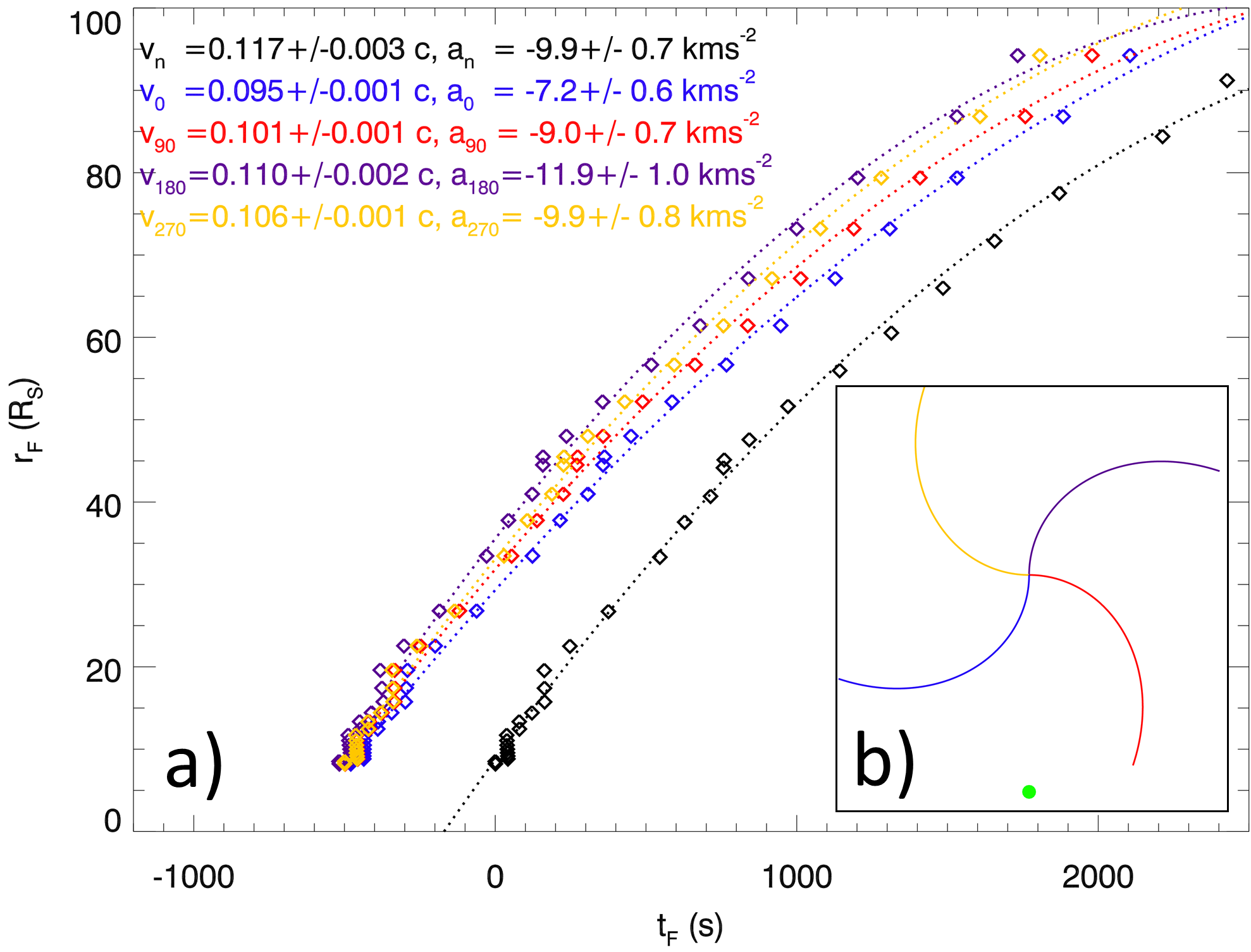}
 \caption{Analysis of measurements recorded by STEREO-A from 14:15 to 16:00 UT on September 27, 2013 assuming the \textit{F-}component:
Panel a): radial distances from the Sun \textit{vs.} uncorrected (in black) and corrected times assuming Parker-spiral propagation
(in blue, red, purple, and orange).
Panel b): Parker-spiral paths and the spacecraft location (in green).}
              \label{fig:spiral}
 \end{figure*}

We also investigated the role of Parker-spiral propagation on estimates of speed and acceleration (Figure \ref{fig:spiral}).
We analyzed the data from STEREO-A  from September 27, 2013 assuming
the \textit{F-}component (Section \ref{S-event}).
The emitting energetic electrons were assumed to travel along four Parker-spiral lines with respect to spacecraft position.
The parameters obtained for various time corrections are not significantly different.
For a comparison, results with no time correction are shown in black.
We thus conclude that the time correction method is applicable to our data set and that it provides us with reasonable results
for electron beam dynamics with typical speeds in the solar wind.

\begin{acknowledgements}
The authors would like to thank the many individuals and institutions who contributed to making
STEREO/\textit{Waves} possible, among which CNES \& CNRS.
This work has been supported by the Praemium Academiae award of The Czech Academy of Sciences.
O.~Kruparova acknowledges the support of the Czech Science Foundation grant GP13-37174P.
J.~Soucek thanks the support of the Czech Science Foundation grant GAP209/12/2394.
O.~Santolik acknowledges additional support from the LH12231 grant.
Financial support by STFC and by the European Commission through
the "Radiosun" (PEOPLE-2011-IRSES-295272) is gratefully acknowledged (E.P.~Kontar).

\end{acknowledgements}


%
%
%
%
%
   %
  %
%
%

\end{document}